\documentclass[aps,showpacs,preprintnumbers,amsmath, amssymb]{revtex4}

\usepackage{float}
\usepackage{graphics,epsfig}
\usepackage{graphicx}
\usepackage{dcolumn}
\usepackage{bm}

\begin{document}
\baselineskip=0.8 cm

\title{{\bf Analytical investigations on non-minimally coupled scalar fields outside neutral reflecting shells}}
\author{Yan Peng$^{1}$\footnote{yanpengphy@163.com}}
\affiliation{\\$^{1}$ School of Mathematical Sciences, Qufu Normal University, Qufu, Shandong 273165, China}

\vspace*{0.2cm}
\begin{abstract}
\baselineskip=0.6 cm
\begin{center}
{\bf Abstract}
\end{center}

We study the existence of scalar fields outside neutral reflecting shells.
We consider static massive scalar fields non-minimally coupled to the
Gauss-Bonnet invariant. We analytically investigated properties
of scalar fields through the scalar field equation.
In the small scalar field mass regime,
we derive a compact resonance formula for
the allowed masses of scalar fields in the composed scalar field and shell
configurations.

\end{abstract}

\pacs{11.25.Tq, 04.70.Bw, 74.20.-z}\maketitle
\newpage
\vspace*{0.2cm}

\section{Introduction}

The famous black hole no hair theorem has attracted a lot of
attention from physicists and mathematicians for decades.
If true, it states that asymptotically flat black holes cannot support
static scalar field hairs outside horizons, for progress see
references \cite{Bekenstein}-\cite{Brihaye} and reviews \cite{Bekenstein-1,CAR}.
This no hair property was usually attributed to the existence
of absorbing boundary conditions at black hole horizons. So it is of great interest to examine whether
such no hair behavior can appear in horizonless spacetimes.

In the horizonless spacetime, Hod recently proved a new type no hair theorem
that static scalar field hairs cannot exist outside asymptotically flat neutral
reflecting stars \cite{Hod-6,Hod-7}.
Furthermore, such no hair theorem also holds for systems constructed with
static scalar fields and neutral horizonless reflecting
shells \cite{Hod-8,Hod-9}. In the asymptotically dS spacetimes, the static scalar field hair
also cannot exist outside neutral horizonless reflecting stars \cite{Bhattacharjee}.
And further studies showed that such no scalar hair theorem is a general property
for horizonless objects with reflecting boundary
conditions \cite{Yan Peng-1}-\cite{Yan Peng-5}.

However, asymptotically flat black holes can support static
scalar fields non-minimally coupled to electromagnetic Maxwell fields,
which violates the spirit of no hair theorems \cite{CARH,PGS}.
Similarly, Hod proved that scalar fields can exist outside
charged horizonless reflecting shells
when considering non-minimally couplings between static scalar fields
and electromagnetic Maxwell fields \cite{SHod}.
In particular, Hod derived a remarkably compact resonance formula for
the allowed masses of the spatially regular scalar
fields supported by charged shell configurations.
In the background of neutral horizonless reflecting shells,
exterior scalar fields also can exist when including non-minimally couplings between scalar fields and
the Gauss-Bonnet invariant \cite{Ypeng}.
Inspired by analysis in \cite{SHod}, we plan to carry out an analytical investigation on configurations
composed of scalar fields and neutral reflecting shells
in the scalar-Gauss-Bonnet gravity.

This work is organized as follows. We firstly introduce the
system with scalar fields non-minimally
coupled to the Gauss-Bonnet invariant outside neutral horizonless reflecting shells.
Then we shall derive a remarkably compact resonance formula for
the allowed masses of scalar fields supported by neutral reflecting shells.
We will summarize our main results in the last section.

\section{Scalar field equations in the reflecting shell background}

We consider scalar fields non-minimally coupled to the Gauss-Bonnet invariant.
The Lagrangian density of the scalar-Gauss-Bonnet gravity is given by \cite{SGB1,SGB2,SGB3,SGB4,SGB5,SGB6,SGB7}
\begin{eqnarray}\label{lagrange-1}
\mathcal{L}=R-|\nabla_{\alpha} \Psi|^{2}-\mu^{2}\Psi^{2}+f(\Psi)\mathcal{R}_{GB}^{2}.
\end{eqnarray}

Here R is the Ricci scalar curvature, $\Psi$ is the scalar field with mass $\mu$
and $f(\Psi)\mathcal{R}_{GB}^{2}$ describes the coupling, where $\mathcal{R}_{GB}^{2}$ is the Gauss-Bonnet invariant \cite{SGB3,SGB4}
\begin{eqnarray}\label{lagrange-1}
\mathcal{R}_{GB}^{2}=R_{\mu\nu\rho\sigma}R^{\mu\nu\rho\sigma}-4R_{\mu\nu}R^{\mu\nu}+R^2
\end{eqnarray}
and $f(\Psi)$ is a general function of $\Psi$.
When neglecting matter fields' backreaction on the metric, the Gauss-Bonnet invariant term is
$\mathcal{R}_{GB}^{2}=\frac{48 M^{2}}{r^{6}}$.
Without loss of generality, the function $f(\Psi)$ can be
putted in a simple form $f(\Psi)=\eta\Psi^2$ in the linear limit,
where $\eta$ is the coupling strength parameter.

The spherically symmetric spacetime is characterized by the
curved line element \cite{SGB4}
\begin{eqnarray}\label{AdSBH}
ds^{2}&=&-g(r)dt^{2}+\frac{dr^{2}}{g(r)}+r^{2}(d\theta^2+sin^{2}\theta d\phi^{2}).
\end{eqnarray}
The radial coordinate depending function $g(r)$ is the metric solution.
The angular coordinates are labeled as $\theta$ and $\phi$ respectively.
And the radius of the shell is defined as $r_{s}$.

With variation methods, we obtain the scalar field equation \cite{SGB1}-\cite{SGB7}
\begin{eqnarray}\label{BHg}
[(\nabla^{\alpha}-i q A^{\alpha})(\nabla_{\alpha}-i q A_{\alpha})-\mu^2+\eta\mathcal{R}_{GB}^{2}]\Psi=0.
\end{eqnarray}

We choose to study a stationary massive
scalar field in the form
\begin{eqnarray}\label{BHg}
\Psi(r,\theta,\phi)=\sum_{lm}\frac{\psi_{lm}(r)}{r}S_{lm}(\theta)e^{im\phi}
\end{eqnarray}
with ${l,m}$ representing the integer harmonic parameters.
$S_{lm}(\theta)$ is the angular scalar eigenfunction with
the eigenvalue $l(l+1)$, where $l$ is the spherical harmonic index \cite{SHod}.
For simplicity, we label the characteristic radial function $\psi_{lm}(r)$ as $\psi(r)$.

The equation (4) and field decomposition (5) yield the ordinary differential equation
\begin{eqnarray}\label{BHg}
[\frac{d^{2}}{dr^{2}}-\mu^{2}-\frac{l(l+1)}{r^{2}}+\frac{48\eta M^{2}}{r^{6}}]\psi=0.
\end{eqnarray}
Here we take the metric outside shells in the flat spacetime limit with $g=1$ and $g'=1$.

At the shell radius, the scalar field satisfies reflecting surface conditions
 \begin{equation}
\psi(r_{s})=0.
\end{equation}

At the infinity, the scalar field is spatially regular,
leading to the vanishing condition
\begin{equation}
\psi(\infty)=0.
\end{equation}

\section{Analytical formula for the allowed masses of scalar fields}

We investigate on properties of scaler fields outside reflecting shells through
the ordinary differential equation (6). The equation can be analytically
studied in two regions $r\ll 1/\mu$ and $r\gg (\sqrt{\eta}M)^{1/2}$.
In the small scalar field mass regime $\mu\ll 1$, we can analyze
the system in the overlapping region $(\sqrt{\eta}M)^{1/2}\ll r\ll 1/\mu$ \cite{SHod}.
In the overlapping region, we apply matching methods to obtain a formula of the
masses of the non-minimally coupled scalar fields outside reflecting shells.

In the limit $r\ll 1/\mu$, the ordinary differential equation (6) can be expressed as
\begin{eqnarray}\label{BHg}
[\frac{d^{2}}{dr^{2}}-\frac{l(l+1)}{r^{2}}+\frac{48\eta M^{2}}{r^{6}}]\psi=0.
\end{eqnarray}

The general mathematical solution of the equation (9) is
\begin{eqnarray}\label{BHg}
\psi(r)=A_{1}r^{\frac{1}{2}}J_{\frac{1}{4}+\frac{1}{2}l}(\frac{2\sqrt{3}\sqrt{\eta}M}{r^{2}})+A_{2}r^{\frac{1}{2}}Y_{\frac{1}{4}+\frac{1}{2}l}(\frac{2\sqrt{3}\sqrt{\eta}M}{r^{2}}),
\end{eqnarray}
where $J_{\nu}(z)$ and $Y_{\nu}(z)$ are Bessel functions of the first and second kinds respectively.
And the coefficients $A_{1}$ and $A_{2}$ are normalization constants.

In the limit $z\rightarrow 0$, the Bessel functions asymptotically behave as s (see Eqs. 9.1.7 and 9.1.9 of \cite{MB})
\begin{eqnarray}\label{BHg}
J_{\nu}(z)=\frac{(z/2)^{\nu}}{\Gamma(1+\nu)}\cdot [1+O(z^{2})],
\end{eqnarray}
\begin{eqnarray}\label{BHg}
Y_{\nu}(z)=-\frac{\Gamma(\nu)}{\pi(z/2)^{\nu}}\cdot [1+O(z^{2})].
\end{eqnarray}

In the radial overlap region $(\sqrt{\eta}M)^{1/2}\ll r\ll 1/\mu$,
there is the relation $\frac{\sqrt{\eta}M}{r^{2}}\ll 1$
and solution (10) of the scalar field equation (9) can be expressed as
\begin{eqnarray}\label{BHg}
\psi(r)=A_{1}\frac{(\sqrt{3}\sqrt{\eta}M)^{\frac{1}{2}l+\frac{1}{4}}}{\Gamma(\frac{1}{2}l+\frac{5}{4})}r^{-l}
-A_{2}\frac{\Gamma(\frac{1}{2}l+\frac{1}{4})}{\pi (\sqrt{3}\sqrt{\eta}M)^{\frac{1}{2}l+\frac{1}{4}}}r^{l+1}.
\end{eqnarray}

In the limit $r\gg (\sqrt{\eta}M)^{1/2}$, the ordinary differential equation (6)
can be approximated by
\begin{eqnarray}\label{BHg}
[\frac{d^{2}}{dr^{2}}-\mu^{2}-\frac{l(l+1)}{r^{2}}]\psi=0.
\end{eqnarray}

The general mathematical solution of (14) can be expressed by
the Bessel functions in the form
\begin{eqnarray}\label{BHg}
\psi(r)=B_{1}r^{\frac{1}{2}}J_{l+\frac{1}{2}}(i\mu r)+B_{2}r^{\frac{1}{2}}Y_{l+\frac{1}{2}}(i\mu r),
\end{eqnarray}
where the coefficients $B_{1}$ and $B_{2}$ are normalization constants.

According to behaviors (11) and (12),
equation (15) can be mathematically expressed as
\begin{eqnarray}\label{BHg}
\psi(r)=B_{1}\frac{(i\mu/2)^{l+\frac{1}{2}}}{\Gamma(l+\frac{3}{2})}\cdot r^{l+1}-B_{2}\frac{\Gamma(l+\frac{1}{2})}{\pi (i\mu/2)^{l+\frac{1}{2}}}\cdot r^{-l}
\end{eqnarray}
for the scalar field in the overlap region $(\sqrt{\eta}M)^{1/2}\ll r\ll 1/\mu$.
Therefore, two analytically derived mathematical expressions (13) and (16) for the scalar field $\psi(r)$ are both valid in the intermediate
radial region
\begin{eqnarray}\label{BHg}
(\sqrt{\eta}M)^{1/2}\ll r\ll 1/\mu.
\end{eqnarray}

This fact allows us to match the solutions (13) and (16) in the region (17)
and obtain relations between coefficients $A_{i}$ and $B_{i}$  as
\begin{eqnarray}\label{BHg}
B_{1}=-A_{2}\frac{\Gamma(\frac{1}{2}l+\frac{1}{4})\Gamma(l+\frac{3}{2})}{\pi}\cdot (-\frac{4}{\sqrt{3}\sqrt{\eta}\mu^{2}M})^{\frac{1}{2}l+\frac{1}{4}},
\end{eqnarray}
\begin{eqnarray}\label{BHg}
B_{2}=-A_{1}\frac{\pi}{\Gamma(\frac{1}{2}l+\frac{5}{4})\Gamma(l+\frac{1}{2})}\cdot (-\frac{\sqrt{3}\sqrt{\eta}\mu^{2}M}{4})^{\frac{1}{2}l+\frac{1}{4}}.
\end{eqnarray}

In the large-argument ($z\rightarrow \infty$), the Bessel functions behave as (see Eqs. 9.2.1 and 9.2.2 of \cite{MB})
\begin{eqnarray}\label{BHg}
J_{\nu}(z)=\sqrt{\frac{2}{\pi z}}\cdot[cos(z-\frac{1}{2}\nu \pi-\frac{1}{4}\pi)]\cdot[1+O(z^{-1})],
\end{eqnarray}
\begin{eqnarray}\label{BHg}
Y_{\nu}(z)=\sqrt{\frac{2}{\pi z}}\cdot[sin(z-\frac{1}{2}\nu \pi-\frac{1}{4}\pi)]\cdot[1+O(z^{-1})],
\end{eqnarray}
and the solution (15) has the asymptotic functional behavior
\begin{eqnarray}\label{BHg}
\psi(r\rightarrow \infty)=B_{1}\sqrt{\frac{2}{i \pi \mu}}cos(i \mu r-\frac{1}{2}l \pi -\frac{1}{2} \pi)+B_{2}\sqrt{\frac{2}{i \pi \mu}}sin(i \mu r-\frac{1}{2}l \pi -\frac{1}{2} \pi).
\end{eqnarray}

The asymptotic behavior (22) and the boundary condition (8) yield the relation
\begin{eqnarray}\label{BHg}
B_{2}=i B_{1}.
\end{eqnarray}

With relations (18), (19) and (23), we derive the relation between $A_{i}$ in the form
\begin{eqnarray}\label{BHg}
(\frac{3\eta \mu^{4} M^{2}}{16})^{\frac{1}{2}l+\frac{1}{4}}=i[\frac{\Gamma(l+\frac{1}{2})\Gamma(l+\frac{3}{2})\Gamma(\frac{1}{2}l+\frac{1}{4})\Gamma(\frac{1}{2}l+\frac{5}{4})}{\pi^{2}}]\frac{A_{2}}{A_{1}}.
\end{eqnarray}

We investigate on properties of scalar fields in the regime $\mu\ll 1$. So there is the relation $\mu r_{s}\ll 1$.
The expression (10) holds at the shell surface $r_{s}$ satisfying $\mu r_{s}\ll 1$.
The boundary condition (7) and the solution (10) yield expression of $A_{2}/A_{1}$ as
\begin{eqnarray}\label{BHg}
\frac{A_{2}}{A_{1}}=-\frac{J_{\frac{1}{4}+\frac{1}{2}l}(\frac{2\sqrt{3}\sqrt{\eta}M}{r_{s}^{2}})}{Y_{\frac{1}{4}+\frac{1}{2}l}(\frac{2\sqrt{3}\sqrt{\eta}M}{r_{s}^{2}})}.
\end{eqnarray}

Substituting the ratio $A_{2}/A_{1}$ of (25) into relation (24), we arrive at the resonance equation
\begin{eqnarray}\label{BHg}
\sqrt{\eta} \mu^{2} M=\frac{4}{\sqrt{3}}(-1)^{\frac{1}{2l+1}}[\frac{\Gamma(l+\frac{1}{2})\Gamma(l+\frac{3}{2})\Gamma(\frac{1}{2}l+\frac{1}{4})\Gamma(\frac{1}{2}l+\frac{5}{4})J_{\frac{1}{4}+\frac{1}{2}l}
(\frac{2\sqrt{3}\sqrt{\eta}M}{r_{s}^{2}})}{\pi^{2}Y_{\frac{1}{4}+\frac{1}{2}l}(\frac{2\sqrt{3}\sqrt{\eta}M}{r_{s}^{2}})}]^{\frac{2}{2l+1}}.
\end{eqnarray}

The relation (17) yields the relation
\begin{eqnarray}\label{BHg}
\sqrt{\eta} \mu^{2} M\ll 1.
\end{eqnarray}

According to equation (26), the relation (27) implies
\begin{eqnarray}\label{BHg}
\frac{2\sqrt{3}\sqrt{\eta}M}{r_{s}^{2}}\thickapprox j_{\frac{1}{4}+\frac{1}{2}l,n},
\end{eqnarray}
where $\{j_{\nu.n}\}_{n=1}^{n=\infty}$ are positive zeros of the Bessel function $J_{\nu}(x)$.

The Bessel functions can be expanded around the positive zeros in the form (see Eq. 9.1.27d of \cite{MB})
\begin{eqnarray}\label{BHg}
J_{\frac{1}{4}+\frac{1}{2}l}(\frac{2\sqrt{3}\sqrt{\eta}M}{r_{s}^{2}})=-J_{\frac{5}{4}+\frac{1}{2}l}(j_{\frac{1}{4}+\frac{1}{2}l,n})\cdot \Delta_{n} \cdot [1+O(\Delta_{n})]
\end{eqnarray}
and
\begin{eqnarray}\label{BHg}
Y_{\frac{1}{4}+\frac{1}{2}l}(\frac{2\sqrt{3}\sqrt{\eta}M}{r_{s}^{2}})=Y_{\frac{1}{4}+\frac{1}{2}l}(j_{\frac{1}{4}+\frac{1}{2}l,n})\cdot [1+O(\Delta_{n})],
\end{eqnarray}
where $\Delta_{n}$ is the small quantity defined as
\begin{eqnarray}\label{BHg}
\Delta_{n}=\frac{2\sqrt{3}\sqrt{\eta}M}{r_{s}^{2}}-j_{\frac{1}{4}+\frac{1}{2}l,n}\ll 1.
\end{eqnarray}

Substituting the expansions (29) and (30) into (26), one obtains the formula
\begin{eqnarray}\label{BHg}
\sqrt{\eta} \mu^{2} M=\frac{4}{\sqrt{3}}(-1)^{\frac{1}{2l+1}}[\frac{\Gamma(l+\frac{1}{2})\Gamma(l+\frac{3}{2})\Gamma(\frac{1}{4}l+\frac{1}{4})\Gamma(\frac{1}{4}l+\frac{5}{4})J_{\frac{5}{4}+\frac{1}{2}l}
(j_{\frac{1}{4}+\frac{1}{2}l,n})}{\pi^{2}Y_{\frac{1}{4}+\frac{1}{2}l}(j_{\frac{1}{4}+\frac{1}{2}l,n})}\Delta_{n}]^{\frac{2}{2l+1}},
\end{eqnarray}
which shows that the allowed values of scalar field masses are discrete.

\section{Conclusions}

We studied the existence of scalar fields outside
neutral reflecting shells in the flat spacetime limit.
We considered static massive scalar fields non-minimally
coupled to the Gauss-Bonnet invariant.
We investigated on properties of scaler fields outside reflecting shells through
the ordinary differential equation (6).
The system is amenable to an analytical analysis
in two regions $r\ll 1/\mu$ and $r\gg (\sqrt{\eta}M)^{1/2}$,
where $\eta$ is the coupling parameter, M is the shell mass
and $\mu$ is the scalar field mass.
In the small scalar field mass regime $\mu\ll 1$, we can analyze
the system in the overlapping region $(\sqrt{\eta}M)^{1/2}\ll r\ll 1/\mu$.
In the overlapping region,
we applied matching methods to derive a remarkably compact resonance formula (32)
for the allowed masses of the supported spatially regular scalar
fields outside neutral horizonless reflecting shells.

\begin{acknowledgments}

This work was supported by the Shandong Provincial Natural Science Foundation of China under Grant
No. ZR2022MA074. This work was supported by a grant from Qufu Normal University
of China under Grant No. xkjjc201906. This work was also supported by the Youth Innovations and Talents Project of Shandong
Provincial Colleges and Universities (Grant no. 201909118).

\end{acknowledgments}

\end{document}